\documentclass[twocolumn,preprintnumbers,amsmath,amssymb]{revtex4}

\usepackage{graphicx}% Include figure files
\usepackage{dcolumn}% Align table columns on decimal point
\usepackage{bm}% bold math
\usepackage{epsf}

\newcommand{\beq}{\begin{equation}}
\newcommand{\eeq}{\end{equation}}
\newcommand{\beqr}{\begin{eqnarray} \nonumber}
\newcommand{\eeqr}{\end{eqnarray}}
\newcommand{\beqrb}{\begin{eqnarray}}
\newcommand{\eeqrb}{\nonumber \end{eqnarray}}

\newcommand{\vect}[1]{\mathbf{#1}}

\newcommand{\fin}{\mbox{ .}}
\newcommand{\coma}{\mbox{ ,}}

\begin{document}

\title{Analytical Study of Diffusive Relativistic Shock Acceleration}

\author{Uri Keshet}

\affiliation{ Institute for Advanced Study, Einstein Drive,
Princeton, NJ 08540, USA }

\altaffiliation{Friends of the Institute for Advanced Study
member}

\email{keshet@sns.ias.edu}

\date{\today}

\begin{abstract}
Particle acceleration in relativistic shocks is studied analytically in
the test-particle, small-angle scattering limit, for an arbitrary
velocity-angle diffusion function $D$. Accurate analytic expressions for
the spectral index $s$ are derived using few ($2-6$) low-order moments of
the shock-frame angular distribution. For isotropic diffusion, previous
results are reproduced and justified. For anisotropic diffusion, $s$ is
shown to be sensitive to $D$, particularly downstream and at certain
angles, and a wide range of $s$ values is attainable. The analysis,
confirmed numerically, can be used to test collisionless shock models and
to observationally constrain $D$. For example, strongly forward- or
backward-enhanced diffusion downstream is ruled out by GRB afterglow
observations.
\end{abstract}

\maketitle

Diffusive (Fermi) shock acceleration (DSA) is believed to be the
mechanism responsible for the production of non-thermal,
high-energy distributions of charged particles in collisionless
shocks in numerous, diverse astronomical systems \cite{FermiAcc}.
Particle acceleration is identified in both non-relativistic and
relativistic shocks, examples of the latter including shocks in
gamma-ray bursts (GRB's, where shock Lorentz factors
$\gamma\gtrsim 100$ are inferred) \cite{grb}, jets of radio
galaxies \cite{Jets}, active galactic nuclei \cite{AGNjets} and
X-ray binaries (micro-quasars) \cite{microquasars}.

Collisionless shocks in general, and the particle acceleration involved in
particular, are mediated by electromagnetic (EM) waves, and are still not
understood from first principles. No present analysis self-consistently
calculates the generation of EM waves and the wave-plasma interactions.
Instead, the particle distribution (PD) $f$ is usually evolved by adopting
some Ansatz for the scattering mechanism (e.g. diffusion in velocity
angle) and neglecting wave generation and shock modification by the
accelerated particles (the "test-particle" approximation). This
phenomenological approach proved successful in accounting for
\emph{non}-relativistic shock observations. For such shocks, DSA predicts
a (momentum $p$) power-law spectrum, $d^3f/dp^3\propto p^{-s}$, where $s$
is a function of the shock compression ratio $r$, $s=3r/(r-1)$
\cite{NonRelDSA}. For strong shocks in an ideal gas of adiabatic index
$\Gamma=5/3$, $s=r=4$ (i.e. $p^2d^3f/dp^3\propto p^{-2}$), in agreement
with observations.

Analysis of GRB afterglow observations suggested that the
ultra-relativistic shocks involved produce high-energy PD's with
$s=4.2\pm0.2$ \cite{grb_s}. This triggered a numerical study of
test-particle DSA in such shocks, $s$ calculated for a wide range of
$\gamma$, various equations of state, and several scattering
mechanisms \cite[][and references therein] {Bednarz98, Kirk00,
Achterberg01}. For isotropic, small-angle scattering in the
ultra-relativistic shock limit, where upstream (downstream) fluid
velocities normalized to the speed of light $c$ approach $\beta_u=1$
($\beta_d =1/3$), spectral indices $s_{0;ur}=4.22\pm0.02$ were found
\cite{Bednarz98, Kirk00, Achterberg01}, in accord with GRB
observations.

DSA analysis is more complicated when the shock is relativistic,
mainly because $f$ becomes anisotropic. Monte-Carlo simulations
\cite{EarlyMonteCarlo, Bednarz98, Achterberg01} and other
numerical techniques \cite{NumericalStudiesA, Kirk00,
NumericalStudiesB} have thus been applied to the problem. For
isotropic, small-angle scattering, an approximate expression for
the spectrum was found \cite{Keshet05},
\begin{equation} \label{eq:s_iso_app}
s_0 = (3\beta_u - 2\beta_u \beta_d^2 +
\beta_d^3) / (\beta_u - \beta_d)\coma
\end{equation}
and shown to agree with numerical results; in particular it yields
$s_{0;ur} = 38/9\simeq 4.22$. Although $s$ was calculated numerically for
various scattering mechanisms \cite{NumericalVeriousDiffusion}, little
qualitative understanding of the dependence of $s$ upon the scattering
Ansatz has emerged. This motivates an analytical study of DSA in
relativistic shocks, which may facilitate/test future self-consistent
shock models.

This letter presents the first rigorous, fully analytical study of DSA in
relativistic shocks. The (common) assumptions made are (i) the
test-particle approximation; (ii) small-angle scattering, given by some
velocity-angle diffusion function $D$; and (iii) $D$ is separable into an
angular part times a space/momentum-dependent part. For discussion and
physical examples of $D$, see \cite{FermiAcc,NumericalStudiesA,Kirk00}. In
the analysis, $N$ angular moments of $f$ are used (and higher moments
neglected) to accurately calculate $s$ even for small $N$. The moments are
generalized in order to accelerate the convergence with $N$ and to obtain
$s$ and $f$ for arbitrary $D$. For isotropic diffusion, the analysis
reproduces and justifies previous results such as
Eq.~(\ref{eq:s_iso_app}), good results obtained even with $N=2$. For
anisotropic diffusion, the dependence of $s$ upon $D$ is analyzed, and is
numerically confirmed and complemented. It is shown that $D$ can be
constrained using observations, e.g. in GRB afterglows. The analysis works
equally well for arbitrarily large $\gamma$, for which numerical methods
become exceedingly difficult. Here we outline the main steps of the
derivation and present its primary implications; the full analysis is
deferred to a detailed, future paper.

\emph{Formalism.---} Consider an infinite shock front at $z=0$, with flow
in the positive z direction. Relativistic particles with momentum
$\vect{\tilde{p}}$ much higher than any characteristic momentum in the
problem are assumed to diffuse in the direction of momentum,
$\tilde{\mu}\equiv \cos (\vect{\tilde{p}}\cdot \vect{\hat{z}}/\tilde{p})$,
according to some diffusion function (DF) $\tilde{D}_{\mu\mu}$. The
steady-state PD then satisfies the stationary transport equation
\cite{NumericalStudiesA}
\begin{equation} \gamma_i(\beta_i+\tilde{\mu}_i)
\frac{\partial f(\tilde{\mu}_i, \tilde{p}_i, z)}{\partial z} =
\frac{\partial}{\partial \tilde{\mu}_i} \left[
\widetilde{D}_{\mu\mu}^{(i)}(\tilde{\mu}_i, \tilde{p}_i, z)
\frac{\partial f}{\partial \tilde{\mu}_i} \right] \coma
\label{eq:transport1}
\end{equation}
where $\gamma=(1-\beta^2)^{-1/2}$ is the fluid Lorentz factor and
$i\in\{u,d\}$ are upstream/downstream indices, written henceforth
only when necessary. Parameters are measured in the fluid frame
(when marked by a tilde) or in the shock frame (otherwise), so the
Lorentz invariant PD $f(\tilde{\mu},\tilde{p},z)$ is the density
in a mixed-frame phase-space. The absence of a characteristic
momentum scale implies that a power-law spectrum is formed, as
verified numerically \cite{Bednarz98, Achterberg01}. With our
assumption that $\widetilde{D}_{\mu\mu}$ is separable in the form
$\widetilde{D}_1(\tilde{\mu}) \widetilde{D}_2(\tilde{p},z)$, we
may thus write Eq.~(\ref{eq:transport1}) as
\begin{equation} (\beta+\tilde{\mu})
\partial_{\tilde{\tau}} q(\tilde{\mu},\tilde{\tau}) = \partial_{\tilde{\mu}}
[(1-\tilde{\mu}^2) \widetilde{D}(\tilde{\mu})
\partial_{\tilde{\mu}} \tilde{q}] \coma
\label{eq:transport2}
\end{equation}
where $f(\tilde{\mu},\tilde{p},z) \equiv
\tilde{q}(\tilde{\mu},\tilde{\tau})\tilde{p}^{-s}$, and the spatial
dimension was rescaled as $\tilde{\tau} \equiv \gamma^{-1} \int_0^z
\widetilde{D}_2(\tilde{p},\check{z}) d\check{z}$. The reduced DF
$\widetilde{D}(\tilde{\mu}) \equiv
\widetilde{D}_1(\tilde{\mu})/(1-\tilde{\mu}^2)$ is constant for
isotropic diffusion.

Continuity across the shock front requires that $f_u(\tilde{\mu}_u,
\tilde{p}_u, z=0) = f_d(\tilde{\mu}_d, \tilde{p}_d, z=0)$, where
upstream and downstream quantities are related by a Lorentz boost of
velocity $c(\beta_u-\beta_d)/(1-\beta_u \beta_d)$. Absence of
accelerated particles far upstream and their diffusion far
downstream imply that $f_u(z \to -\infty) = 0$ and
$f_d(\tilde{\mu}_d,\tilde{p}_d,z \to +\infty) = \tilde{f}_\infty
\cdot \tilde{p}_d^{-s}$, where $\tilde{f}_\infty>0$ is constant
(isotropic PD). Eq.~(\ref{eq:transport2}) has been solved
numerically under the above boundary conditions for specific choices
of $\tilde{D}(\mu)$ \cite{Kirk00}.

\emph{Angular Moments.---} In the shock frame, only a small fraction of
the particles travel nearly parallel to the flow ($\mu\simeq\pm 1$),
unlike the upstream frame where a substantial part of $f$ is beamed into a
narrow, $\tilde{\mu}_u <-1+ \gamma_u^{-1}$ cone. Angular moments of the
shock-frame PD, $\int_{-1}^{1} \mu^n f\,d\mu$, thus diminish as $n\geq0$
increases, suggesting that the problem can be approximately formulated
using a finite number of low-$n$ moments. Next, we derive and solve
equations for the spatial behavior of these moments; the boundary
conditions are then used to determine $s$.

We work in the shock frame and use only shock frame variables
henceforth, incorporating the Lorentz boosts into the transport
equations. Writing Eq.~(\ref{eq:transport2}) in the shock frame,
using $p=\tilde{\gamma_i}\tilde{p}_i(1+\beta_i\tilde{\mu}_i)$ and
$\mu=(\tilde{\mu}_i+\beta_i)/(1+\beta_i\tilde{\mu}_i)$, we find
that on each side of the shock
\begin{equation}
\mu \, \partial_\tau q(\mu,\tau) = \frac{ \partial_{\mu}\left\{
(1-\mu^{2})D(\mu)\partial_{\mu}\left[(1-\beta\mu)^{s}q\right]\right\}
}{(1-\beta\mu)^{s-3}} \label{eq:transport_shock_frame} \coma
\end{equation}
where we defined $q(\mu,\tau)p^{-s} \equiv
\tilde{q}(\tilde{\mu},\tilde{\tau})\tilde{p}^{-s}$, $D(\mu) \equiv
\widetilde{D}(\tilde{\mu})$, and $\tau\equiv \gamma^4 \tilde{\tau}$.
An immediate consequence is that
\begin{equation}  \label{eq:flux_conservation}
\int_{-1}^{1} (1-\beta\mu)^{s-3} \mu
\,q(\mu,\tau)d\mu=g=\mbox{const} \coma
\end{equation}
where the boundary conditions yield $g_u=0$ and $g_d\propto
\tilde{f}_{\infty}$. Eq.~(\ref{eq:flux_conservation}) is an
expression of particle number and energy conservation in the fluid
frame \cite{Keshet05}. A more general corollary of
Eq.~(\ref{eq:transport_shock_frame}) is that for any function
$h(\mu)$ for which $h/\mu$ is finite and continuously
differentiable,
\begin{align} \label{eq:corollary1}
& \partial_{\tau} \int_{-1}^{1} h(\mu) q(\mu,\tau) d\mu =
\int_{-1}^{1} (1-\beta\mu)^{s}q \\
& \times \partial_{\mu} \left\{
\left(1-\mu^{2}\right)D(\mu)\partial_{\mu} \left[ h(\mu)
\mu^{-1}\left(1-\beta\mu\right)^{-(s-3)}\right] \right\} \nonumber
d\mu.
\end{align}

The spatial dependence of an angular moment $F_n(\tau) \equiv
\int_{-1}^{1} \mu^n q(\mu,\tau)\,d\mu$ is given by
Eq.~(\ref{eq:corollary1}) with $h=\mu^n$, if $n\geq 1$. Expanding
the RHS integrand as a power series around $\mu=0$, we find that
$\partial_\tau F_n$ is given by a linear combination of the moments
$F_{a}, F_{a+1} \ldots F_{b}$. Here,
$a=\mbox{Max}\left\{0,n-3\right\}$, $b=n+2+n_D$, and $n_D$ is
defined as the order of the power series of $D$ around $\mu=0$,
$D(\mu)\propto 1+d_1 \mu+d_2\mu^2+\cdots$, such that $n_D=0$ for
isotropic diffusion. In matrix form, we may write
\begin{equation}
\label{eq_moments_ODEs}
\partial_{\tau}\mathbf{F}(\tau)=\mathbf{A}\cdot\left(\begin{array}{c}
F_{0}(\tau) \\ \mathbf{F}(\tau) \end{array}\right),
\end{equation}
where $\mathbf{F}\equiv \left(F_{1},F_{2},\ldots\right)^{T}$. The
(infinite) matrix $\mathbf{A}$ depends on $\beta$, $s$ and
$D(\mu)$, so $\mathbf{A}_u\neq \mathbf{A}_d$. The boundary
conditions imply that $\{F_n\}$ are continuous across the shock
front $\tau=0$, vanish as $\tau \rightarrow -\infty$, and are
finite for $\tau \rightarrow +\infty$. We show that a good,
converging (in $N$) approximation to the solution is obtained
using a small number $N+1$ of moments, $F_0\ldots F_N$, and
neglecting higher order terms. This reduces $\mathbf{F}$ to an
$N$-component vector and $\mathbf{A}$ to an $N\times(N+1)$ matrix.

The spatial dependence of the zero'th moment $F_0$ cannot be related to
$n\geq0$ angular moments through Eq.~(\ref{eq:corollary1}). This
difficulty can be circumvented by adding an additional constraint to the
system of equations. Expanding the integrand in
Eq.~(\ref{eq:flux_conservation}) yields $\mathbf{G}(\beta,s)\cdot
\mathbf{F}(\tau)=g$, or
\begin{equation} \label{eq:Flux_conservation_moments}
F_{1}-(s-3)\beta
F_{2}+\frac{1}{2}(s-3)(s-4)\beta^{2}F_{3}+\cdots=g\coma
\end{equation}
independent of $D(\mu)$. Eq.~(\ref{eq:Flux_conservation_moments})
and its spatial derivative can be used to eliminate $F_0$ and $F_N$
from Eq.~(\ref{eq_moments_ODEs}), giving an approximate, closed set
of $\check{N}= N-1$ (or less, if $A$ is degenerate) ordinary
differential equations (ODE's),
\begin{equation} \label{eq:reduced_ODEs}
\partial_{\tau} \mathbf{\check{F}}(\tau) = \mathbf{\check{A}}
\cdot \mathbf{\check{F}} + g\mathbf{\check{G}} \coma
\end{equation}
where $\mathbf{\check{F}} \equiv
(F_1,F_2,\ldots,F_{\check{N}})^T$. Here, $\mathbf{\check{A}}\in
M_{\check{N}}$ and $\mathbf{\check{G}}$ are functions of
$\mathbf{A}$ and $\mathbf{G}$ obtained by combining
Eqs.~(\ref{eq_moments_ODEs}) and
(\ref{eq:Flux_conservation_moments}). For the relevant range of
$s$, $\mathbf{\check{A}}$ is diagonalizable over $\mathcal{R}$.
Thus, there exists a real matrix $\mathbf{P}\in M_{\check{N}}$,
such that $(\mathbf{P}^{-1}\mathbf{\check{A}}\mathbf{P})_{mn} =
\lambda_n \delta_{mn}$ and the eigenvalues $\{\lambda_n\}$ of
$\mathbf{\check{A}}$ are real. The solution of
Eq.~(\ref{eq:reduced_ODEs}) is then
\begin{equation} \label{eq:ODEs_solutions}
F_n(\tau) = \sum_{m=1}^{\check{N}} \left[ P_{nm}c_m e^{\lambda_m
\tau} - gP_{nm}(\mathbf{P}^{-1}\mathbf{\check{G}})_m
\lambda_m^{-1} \right].
\end{equation}

The integration constants $c_{m,u}$ and $c_{m,d}$ along with $g_d$
and $s$ constitute $2\check{N}+1$ free parameters, as $\mathbf{F}$
is unique only up to an overall constant. The boundary conditions
typically impose $2\check{N}+2$ constraints, so the problem is
over-constrained. It is natural to relax the requirement that $F_N$
be continuous across the shock, as it is relatively small and
$\partial_\tau F_N$ is relatively sensitive to neglected moments
$n\gtrsim N$. The resulting discontinuity of $F_N$ at $\tau=0$ then
provides an estimate of the approximation accuracy.

\begin{figure}[h]
\centerline{\epsfxsize=9cm \epsfbox{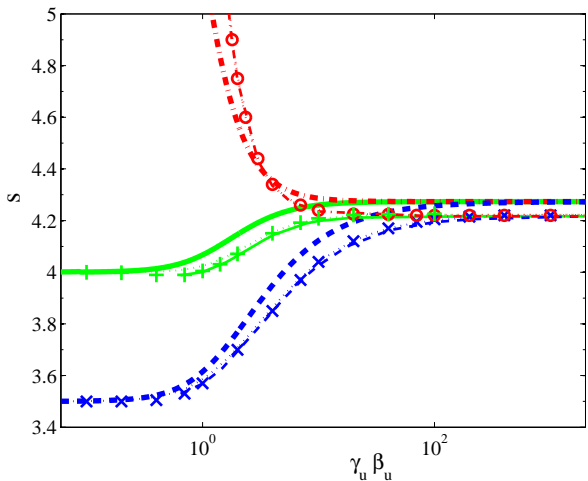}}
\caption{\label{fig:SIso} DSA for isotropic diffusion according to
analytic approximations involving $F_0\ldots F_2$
[Eq.~(\ref{eq_s_approx_F2}), heavy curves] or $L_0\ldots L_5$
(light curves), numerical calculations (Ref. \citep{Kirk00},
symbols) and Eq.~(\ref{eq:s_iso_app}) (dotted curves). Three types
of strong shocks (see \cite{KirkDuffy99}) are examined: using the
J\"{u}ttner-Synge equation of state (solid curves/crosses),
assuming a fixed adiabatic index $\Gamma=4/3$ (dashed
curves/x-marks), and assuming a relativistic gas where $\beta_u
\beta_d=1/3$ (dash-dotted curves/circles).  }
\end{figure}

As an example, consider the simplest non-trivial approximation,
where moments $F_n$ with $n>N=2$ are neglected. This is motivated,
for example, by the suppression of $G_n$ [see
Eq.~(\ref{eq:Flux_conservation_moments})], $A_{1,n}$ and $A_{2,n}$
for $n>2$, if $s\simeq 4$. Here $\check{N}=1$, and the (single) ODE
yields $F_1(\tau)= c e^{\lambda \tau} + g \check{G} \lambda^{-1}$.
For isotropic diffusion, $\lambda = 6\beta -(s-2)(s-3)(s-4)\beta^3$.
In most cases of interest (where $s<5$) $\lambda>0$, corresponding
to exponential decay upstream and divergence downstream. Therefor,
$c_{d}$ must vanish, so $f_d$ is uniform in this approximation. This
gives a cubic equation for $s$, independent of $g_d,c_u$,
\begin{equation} \label{eq_s_approx_F2}
(s-3)\beta_{u} = \left. \frac{F_{1}}{F_{2}} \right|_{\tau=0} =
\beta_{d}+\frac{2(s-1)\beta_{d}}{2+(s-2)(s-3)\beta_{d}^{2}} \fin
\end{equation}
The real root of this equation, shown in Figure \ref{fig:SIso},
already agrees with numerical results and with
Eq.~(\ref{eq:s_iso_app}) at a $\sim 5\%$ level \footnote{With
respect to the extreme case $s(\Gamma=1)=3$ \cite{Keshet05}.}; in
particular it yields $s_{0;ur}=4.27$. However, in this simple
approximation $F_0$ is not continuous across the shock front (it has
a $\lesssim20\%$ jump), $s$ does not depend on $D_u(\mu)$, and its
dependence on $D_d(\mu)$ is inaccurate.

The convergence of the approximation with N depends on $D$, more
moments required in general as $n_D$ and $|d_n|$ increase. Due to
degeneracies in $\mathbf{A}$, $N=2,3,4,5,6,\ldots$ correspond to
$\check{N}=1,1,2,3,3,\ldots$, and $N\geq 6$ is required in order
to improve the $N=2$ approximation, e.g. $s_{0;ur}(N=6)=4.24$. The
convergence is significantly accelerated if the moments are
properly generalized.

\emph{Generalized Moments.---} The analysis can be generalized in
many ways by exploiting the freedom in the definition of the
angular moments. For example, if we define $F_n(\tau)\equiv
\int_{-1}^1 q(\mu,\tau) w_n(\mu) f_n(\mu) \,d\mu$, where $\{f_n\}$
is an orthonormal basis in the interval $-1\leq\mu\leq 1$ with
weight functions $\{w_n\}$, then the moments $\{F_n\}$ are simply
the expansion coefficients of $q$ in terms of the basis $\{f_n\}$,
i.e. $q(\mu, \tau) = \Sigma_n F_n(\tau) f_n(\mu)$ \footnote{Note
that taking the (numerically calculated) eigenfunctions of Eq.
(\ref{eq:transport_shock_frame}) as $\{f_n\}$, and $w_n=\mu$, is
analogous to \cite{Kirk00}.}. With this choice of $\{F_n\}$, the
analysis provides direct information about the PD $f$. Additional
constraints on $f$ \cite[e.g.][]{Keshet05} may thus be used to
estimate/improve the accuracy of the approximation. The analysis
is mostly unchanged; it remains analytically tractable as long as
$\mathbf{A},\mathbf{G}$ can be determined analytically, although
the analogue of Eq.~(\ref{eq_s_approx_F2}) is in general no longer
a polynomial, becoming a transcendental equation in $s$.

As an example, let $f_n(\mu)=[(2n+1)/2]^{1/2}P_n(\mu)$ and
$w_n=1$, where $P_n(\mu)$ is the Legendre polynomial of order $n$.
The moments $\{F_n\}$ thus equal the coefficients $\{L_n\}$ of the
Legendre series of $q$. Here we cannot take $N=2$, because the
corresponding eigenvalue $\lambda$ is negative. For
$N=3,4,5,\ldots$ we find $\check{N}=2,3,4,\ldots$, with
$\check{N}_u=1,2,2,\ldots$ ($\check{N}_d=1,1,2,\ldots)$
non-vanishing exponents upstream (downstream). The resulting
spectrum for isotropic diffusion is accurate to $\sim 5\%$, $2\%,
1\%,\ldots$; in particular $s_{0;ur}=4.16, 4.21, 4.22, \ldots$.
These results illustrate the rapid convergence of the
Legendre-based analysis. The method is equally applicable for
arbitrarily high $\gamma$, whereas numerical methods become
exceedingly difficult as the ultra-relativistic limit is
approached.

\begin{figure}[h]
\centerline{\epsfxsize=9cm \epsfbox{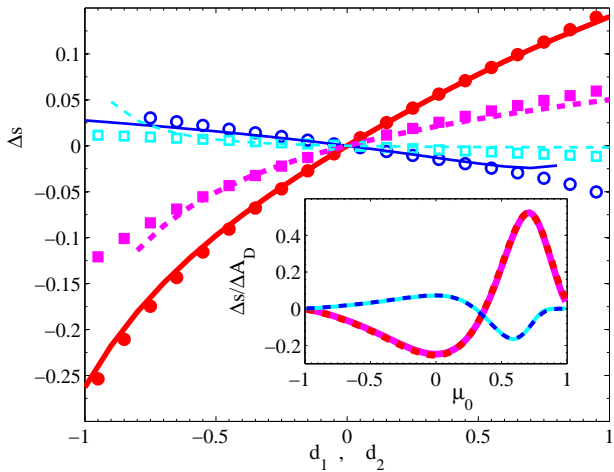}}
\caption{\label{fig:SNonIso} DSA for anisotropic diffusion. The
spectrum is calculated numerically/with $L0\ldots L5$
(symbols/curves) for $D=1+d_n \mu^n$ in a $\gamma=10$, strong shock
with the J\"{u}ttner-Synge equation of state. Plotted is $\Delta
s=s-s_0$ vs. $d_n$ for $n=1$ (circles/solid curves) and $n=2$
(squares/dashed curves), downstream/upstream (filled/open symbols,
heavy/light curves). Inset: local deviations from an isotropic DF
down/upstream (heavy/light curves): $D=1+d_0 \Theta
(w_0-|\mu-\mu_0|)$ (solid curves) and $D=1+d_0 \exp [-(\mu-\mu_0)^2
/ 2w_0^2 ]$ (dashed curves), where $(d_0,w_0) = (0.1, 0.05)$ and
$\Theta$ is the Heaviside step function. }
\end{figure}

Isotropic (anisotropic) diffusion results for $s$ with $N=5$
Legendre moments are shown in Figure \ref{fig:SIso} (Figure
\ref{fig:SNonIso}). For isotropic diffusion, previous results are
reproduced, lending rigor support for Eq.~(\ref{eq:s_iso_app}). The
lowest order moments converge quickly, giving a rough estimate of
$q(\mu,\tau)$ as a sum of components exponentially decaying in
$|\tau|$ and, in the downstream, also a uniform part. The
analytically calculated $s,q$ compare favorably with numerical
results, which we calculate in the eigenfunction method of
\cite{Kirk00}.

\emph{Anisotropic diffusion.---} Qualitatively good results for
$d_{n\leq3}\neq 0$; $d_{n\geq 4}=0$ are already obtained with $N=3$
Legendre-based moments. More complicated forms of the DF, involving
additional $d_n\neq 0$ terms, require progressively larger $N$.
Figure \ref{fig:SNonIso} shows that large $\Delta s=s-s_0$ values
can be obtained for anisotropic diffusion in relativistic shocks.
The spectrum is more sensitive to $D_d$ than it is to $D_u$, by a
factor of a few. It is more sensitive to $d_1$ than to $d_2$;
$|\Delta s(D=1+d_n \mu^n)|$ is roughly monotonically decreasing with
$n$. A linear, forward- (backward-) enhanced downstream DF,
$D_d=1+d_1 \mu$, yields $s>4.3$ ($s<4.0$) in a $\gamma>10$ shock, if
$d_1>0.6$ ($d_1<-0.85$). Some forward- (backward-) enhanced choices
of $D$ lead to more extreme spectra, e.g. $s_{ur}>4.6$
($s_{ur}<3.7$). Constraints on $D_d$ may thus be imposed using
observations that rule out such $s$ values, e.g. in GRB afterglows.

In order to elucidate the role of $D$, we numerically calculate $s$
for small deviations $\Delta D$, localized around some angle
$\mu_0$, from an otherwise isotropic DF. The parameter $\Delta s /
\Delta A_D$, where $\Delta A_D \equiv \int_{-1}^1 \Delta D\,d\mu$,
depends only weakly on the exact form of such $\Delta D$. Figure
\ref{fig:SNonIso} (inset) shows that $s$ is sensitive to $\mu_0$,
especially downstream. In general, the spectrum hardens when $D_d$
is enhanced at $\mu<\mu_c\simeq 0.4$ or diminished at $\mu>\mu_c$,
and vice versa. Angles for which $\mu<\mu_c$ ($\mu>\mu_c$) roughly
correspond to negative (positive) flux in momentum space,
$j_\mu\propto -D(\mu)(1-\mu^2)\partial_\mu q$, qualitatively
accounting for this behavior. One way to see this is to note that
the spectrum hardens when the average fractional energy gain $f_{E}$
(of a particle crossing, say, to the downstream and returning
upstream) or the return probability (to the upstream) $P_{ret}$
increase; $s=3+\ln(1/P_{ret})/\ln f_{E}$ \cite{Achterberg01}. For
example, enhanced diffusion in angles where $j_\mu<0$ shifts $q_d$
towards the upstream, raising both $P_{ret}$ and $f_E$. The above
trend is roughly reversed upstream, but $\Delta s$ is smaller there
and more sensitive to details of $\Delta D_u$.

I thank E. Waxman and D. Vinkovi{\' c} for fruitful discussions.


\begin{thebibliography}{99}

\bibitem{FermiAcc} For reviews see
R. Blandford \& D. Eichler, Phys. Rep. {\bf 154}, 1 (1987); M. A.
Malkov. \& L. O'C. Drury, Rep. Prog. Phys. {\bf 64}, 429 (2001).

\bibitem{grb}  T. Piran, Phys. Rep. {\bf 333}, 529 (2000); P. M\'esz\'aros, ARA\&A {\bf 40}, 137 (2002);
E. Waxman, Lect. Notes Phys. {\bf 598}, 393 (2003).

\bibitem{Jets} M. C. Begelman, M. J. Rees, \& M. Sikora, Astrophys. J. {\bf 429}, L57 (1994);
L. Maraschi, in AGNs: from Central Engine to Host Galaxy, Eds. S.
Collin, F. Combes \& I. Shlosman. ASP, Conference Series, {\bf
290} 275 (2003).

\bibitem{AGNjets} M. J. Rees, Nature {\bf 229}, 312 (1971);
M. C. Begelman, R. D. Blandford \& M. J. Rees, Rev. Mod. Phys.
{\bf 56}, 255 (1984); R. A. Laing, in Astrophysical Jets, ed. D.
Burgarella, M. Livio \& C. P. O'Dea (Cambridge: Cambridge Univ.
Press), 95 (1993).

\bibitem{microquasars} For review see R. Fender, astro-ph/0303339 [in Compact Stellar X-Ray Sources,
edited by W.H.G. Lewin and M. van der Klis (Cambridge University
Press, Cambridge, to be published)].

\bibitem{NonRelDSA}
G. F. Krymskii, Dokl. Akad. Nauk SSSR, {\bf 234}, 1306 (1977); W.
I. Axford, E. Leer \& G. Skadron, Proc. 15th Int. Cosmic Ray
Conf., Plovdiv (Budapest: Central Research Institute for Physics)
{\bf 11}, 132 (1977); A. R. Bell, Mon. Not. R. Astron. Soc. {\bf
182}, 147 (1978); R. D. Blandford \& J. Ostriker, Astrophys. J.
{\bf 221}, L29 (1978).


\bibitem{grb_s} E. Waxman, Astrophys. J. {\bf 485}, L5 (1997);
D. L. Freedman \& E. Waxman, Astrophys. J. {\bf 547}, 922 (2001);
I. Berger, S. R. Kulkarni \& D. A. Frail, Astrophys. J. {\bf 590}, 379 (2003).

\bibitem{Bednarz98} J. Bednarz \& M. Ostrowski, Phys. Rev. Lett. {\bf 80}, 3911
(1998).

\bibitem{Kirk00} J. G. Kirk, A. W. Guthmann, Y. A. Gallant, \& A. Achterberg, Phys. Rev. {\bf 542}, 235
(2000).

\bibitem{Achterberg01} A. Achterberg, Y. A. Gallant, J. G. Kirk, \& A. W. Guthmann, Mon. Not. R. Astron. Soc. {\bf 328},
393 (2001).

\bibitem{EarlyMonteCarlo}
K.~R. {Ballard}, \& A.~F. {Heavens}, Mon. Not. R. Astron. Soc.
{\bf 259}, 89 (1992); M. {Ostrowski}, Mon. Not. R. Astron. Soc.,
{\bf 264}, 248 (1993).

\bibitem{NumericalStudiesA}
J. G. Kirk \& P. Schneider, Astrophys. J. {\bf 315}, 425 (1987);
A. F. Heavens \& L. O'C Drury, Mon. Not. R. Astron. Soc. {\bf
235}, 997 (1988).

\bibitem{NumericalStudiesB}
M. Vietri, Astrophys. J. {\bf 591}, 954 (2003); P. {Blasi} \& M.
{Vietri}, Astrophys. J., {\bf 626}, 877 (2005).

\bibitem{Keshet05} U. Keshet \& E. Waxman, Phys. Rev. Lett. {\bf
94}, 111102 (2005).

\bibitem{NumericalVeriousDiffusion}
D.~C. {Ellison} \& G.~P. {Double}, Astroparticle Phys., {\bf 18},
213 (2002); M. {Lemoine} \& G. {Pelletier}, Astrophys. J. Lett.,
{\bf 589}, L73 (2003); A. Meli \& J. J. Quenby, Astroparticle
Phys., {\bf 19}, 649 (2003); J. {Niemiec} \& M. {Ostrowski},
Astrophys. J., {\bf 610}, 851 (2004); D.~C. {Ellison} \& G.~P.
{Double}, Astroparticle Phys., {\bf 22}, 323 (2004); J. {Bednarz},
PASJ, {\bf 56}, 923 (2004); M. {Lemoine} \& B. {Revenu}, Mon. Not.
R. Astron. Soc., {\bf 366}, 635 (2006).

\bibitem{KirkDuffy99} J. G. Kirk \& P. Duffy, J. Phys. G: Nucl. Part. Phys. {\bf 25}, R163
(1999).


\end{thebibliography}
\end{document}